\documentclass[reprint,amsmath,amssymb,aps,]{revtex4-1}
\usepackage{graphicx}
\usepackage{dcolumn}
\usepackage{bm}
\usepackage{bbm}
\usepackage{enumerate}
\usepackage{color}
\begin{document}
\title{Game theory, statistical mechanics, \\and income inequality}
\author{Venkat Venkatasubramanian}
 \email{To whom correspondence should be addressed. venkat@columbia.edu}
 \altaffiliation{Department of Industrial Engineering and Operations Research, Columbia University.}
 \altaffiliation{Department of Computer Science, Columbia University.}
\author{Yu Luo}
 \email{yl2750@columbia.edu}
\affiliation{Department of Chemical Engineering, Columbia University}
\author{Jay Sethuraman}
\email{jay@ieor.columbia.edu}
 \affiliation{Department of Industrial Engineering and Operations Research, Columbia University}\date{\today}
\begin{abstract}
The widening inequality in income distribution in recent years, and the associated excessive pay packages of CEOs in the U.S. and elsewhere, is of growing concern among policy makers as well as the common person. However, there seems to be no satisfactory answer, in conventional economic theories and models, to the fundamental question of what kind of pay distribution we ought to see, at least under ideal conditions, in a free market environment and whether this distribution is fair. We propose a game theoretic framework that addresses these questions and show that the lognormal distribution is the fairest inequality of pay in an organization comprising of homogenous agents, achieved at equilibrium, under ideal free market conditions. We also show that for a population of two different classes of agents, the final distribution is a combination of two different lognormal distributions where one of them, corresponding to the top $\sim$3-5\% of the population, can be misidentified as a Pareto distribution. Our theory also shows the deep and direct connection between potential game theory and statistical mechanics through entropy, which is a measure of fairness in a distribution. This leads us to propose the {\em fair market hypothesis}, that the self-organizing dynamics of the ideal free market, i.e.,~Adam Smith's ``invisible hand'', not only promotes efficiency but also maximizes fairness under the given constraints. 
\end{abstract}
\maketitle

\section{Introduction}

In recent years, there has been growing concern over the widening inequalities in income and wealth distributions in the U.S. and elsewhere~\cite{krugman2002for,stiglitz2012price,piketty2014capital}. The statistics are troubling~--~for instance, as of 2010, the top 1\% of households in the U.S. owned 35.4\% of all privately held wealth~\cite{domhoff2005power}, and it had risen from a low of about 20\% in 1976.

An important source of the wealth inequality is a similar trend in the income and pay or wage distributions. Income remains highly concentrated, with the top 1\% of income earners received 17.2\% of all income in 2009, and that's up from 12.8\%  in 1982~\cite{piketty2003income,domhoff2005power}. A related trend of equally great concern is the runaway pay packages for CEOs which are reflected in the extraordinarily high CEO pay ratios in the U.S.~\cite{anderson2008executive,hargreaves2014can}.  There is much discussion both in academic literature and popular press about what all these mean, what the consequences are, and what can or should be done about it~\cite{bebchuck2004pay,mishel2005state,jones2009who,anderson2013executive,kristof2014were}.

Obviously, before any policy actions, if any, are taken to address these challenges, we need to understand more deeply why and how such  inequalities occur. Since different people have different abilities and therefore make different contributions, we do expect to see unequal distributions in income and wealth. So, a certain level of  inequality is to be expected. But, at the risk of sounding oxymoronic,  what is the fairest inequality? In particular, in an ideal free market  environment , what is the inequality we ought to see? 
While there is extensive empirical literature on income and wealth distributions, and we cite only a sample here~\cite{anderson2008executive,mishel2005state,kato2006ceo,bebchuck2004pay,champernowne1953model,champernowne1998economic,willis2004evidence}, there is no satisfactory answer to such questions in conventional economic theories~\cite{chakrabarti2013econophysics}. Instead of just relying on empirical data alone, can we predict, at least under ideal conditions, what to expect from a theoretical analysis? Two fundamental questions one would like answered are: {\em What kind of pay distribution will arise, under ideal conditions, in a free market environment comprising of utility maximizing employees and profit maximizing companies? Is this distribution fair?} The answers to these questions can serve as a fundamental benchmark against which we can evaluate the distributions seen in real life. This reference can help us measure and understand the deviations caused by non-idealities under actual conditions, and to develop appropriate policy frameworks and incentive structures to try to correct the inequalities. It can give us a quantitative  basis for understanding and developing pay packages for executives, tax policies, etc.

In  the past decade or so, there has been much work in the econophysics community to model income and wealth distributions by applying concepts and techniques from statistical mechanics~\cite{willis2004evidence,stanley1999econophysics,dragulescu2000statistical,yakovenko2009colloquium,yakovenko2012statistical,richmond2006review,chatterjee2007economic,levy1996power,bouchaud2000wealth,souma2001universal,chatterjee2005econophysics,smith2008classical,willis2011wealth,chakrabarti2013econophysics,cho2014physicists,axtell2014endogenous}. While these models are quite interesting and instructive, they haven't, however,  bridged the rather wide conceptual gulf that exists between economics and econophysics~\cite{gallegati2006worrying,ormerod2010current}, particularly in two crucial areas. One, the typical particle model of agent behavior in econophysics assumes agents to have nearly ``zero intelligence'', acting at random, with no intent or purpose. This does not sit well with an extensive body of economic literature spanning several decades,  where one models, in the ideal case, a perfectly rational agent whose goal is to maximize its utility or profit by acting strategically, not randomly. From the perspective of an economist, it is quite reasonable to ask ``How can theories and models based on the collective behavior of purpose-free, random, molecules explain the collective behavior of goal-driven, optimizing, strategizing men and women?''

Another conceptual stumbling block is the role of entropy in economics. In statistical thermodynamics, equilibrium is reached when entropy, which is a measure of randomness or uncertainty, is maximized. So, an economist wonders, why would maximizing randomness or uncertainty be helpful in economic systems? We all know that markets are stable, and function well, when things are orderly, with less uncertainty, not more. As Amartya Sen observed~\cite{sen1997economic}, ``Given the association of doom with entropy in the context of thermodynamics it may take a little time to get used to entropy as a good thing (`How grand, entropy is on the increase!'), \dots'' Similar objections were raised by Paul Samuelson~\cite{samuelson1990gibbs}: ``As will become apparent, I have limited tolerance for the perpetual attempts to fabricate for economics concepts of `entropy' imported from the physical sciences or constructed by analogy to Clausius-Boltzmann magnitudes.'' Thus, we run into major conceptual hurdles in the typical statistical mechanics-based approaches to problems in economics, particularly  in the study of pay, income and wealth distributions.

Besides these conceptual challenges, there is also a technical one due to the nature of the datasets in economics. As Ormerod~\cite{ormerod2010current} and Perline~\cite{perline2005strong} discuss, one can easily misinterpret data from lognormal distributions, particularly from truncated datasets, as inverse power law or other distributions. Therefore, empirical verification of econophysics models is still in the early stages.

Addressing one of the two conceptual challenges, Venkatasubramanian proposed an  information-theoretic framework \cite{venkatasubramanian2009what,venkatasubramanian2010fairness} wherein he identified that entropy really is a measure of fairness in a distribution, not just randomness or uncertainty, which then makes it an appropriate candidate in economics. In this paper, we follow up on this line of enquiry and address the other critical challenge of reconciling the behavior of goal-driven, teleological, agents with that of purpose-free, randomly driven molecules. We start from a familiar ground in economics, namely, game theory, to develop a new conceptual framework to address the pay distribution questions we raised above. This leads to surprising and useful insights about a deep connection between game theory and statistical mechanics, paving the way for a general theoretical framework that unifies the dynamics of purposeful animate agents with that of purpose-free inanimate ones.

\section{Pay Distribution in an Ideal Free Market Environment: Formulating the Problem}

We follow Venkatasubramanian's~\cite{venkatasubramanian2010fairness} approach in formulating the problem and restate it here for the convenience of the reader. Consider a competitive, dynamic, free market environment comprising of a large number of utility maximizing rational agents as employees and profit maximizing rational agents as corporations. We assume an ideal environment where the market is perfectly competitive, transaction costs are negligible, and no externalities are present. In this ideal free market, employees are free to switch jobs and move between companies in search of better utilities.  Similarly, companies are free to fire and hire employees in order to maximize their profits. We do not consider the effect of taxes.

We also assume that a company needs to retain all its employees in order to survive in this competitive market environment. Thus, a company will take whatever steps necessary, allowed by its constraints, to retain all its employees. Similarly, all employees need a utility to survive and that they will do whatever is necessary, allowed by certain norms, to stay employed. We assume that neither the companies nor the employees engage in illegal practices such as fraud, collusion, and so on.

\begin{figure}[h]
\centerline{\includegraphics[width=1\linewidth]{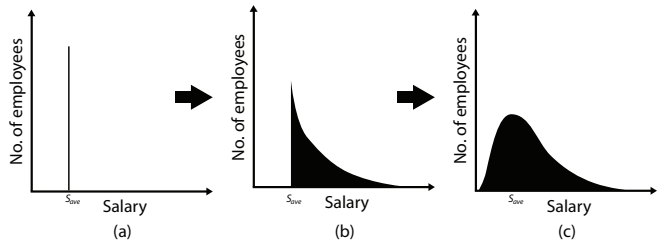}}
\caption{Spreading of the salary distribution under competition in a free market environment}\label{spread}
\end{figure}

In this ideal free market, consider a company A with $N$ employees and a salary budget of $M$, with an average salary of $S_\text{ave} = M/N$. Let us assume that there are $n$ categories of employees~--~ranging from secretaries to the CEO, contributing in different ways towards the company's overall success and value creation. All employees in category $i$ contribute value $V_i, i \in \{1,2,\dots, n\}$, such that $V_1 < V_2 <\dots< V_n$. Let the corresponding value at $S_\text{ave}$ be $V_\text{ave}$, occurring at category $s$. Since all employees are contributing unequally, some more some less, they all need to be compensated differently, commensurate with their relative contributions towards the overall value created by the company. Instead, A has an egalitarian policy that all employees are equal and therefore pays all of them the same salary, $S_\text{ave}$, irrespective of their contributions. The salary of the CEO is the same as that of an administrative assistant in the mail room. This salary distribution is a sharp vertical line at $S_\text{ave}$, as seen in figure~\ref{spread}(a), a Kronecker delta function. 
As noted, while this may seem fair in a social or moral justice sense, clearly it is not in an economic sense. If this were to be the only company in the economic system, or if A is completely isolated from other companies in the economic environment, the employees will be forced to continue to work under these conditions as there is no other choice.

However, in an ideal free market system there are other choices. Therefore, all those employees who contribute more than the average -- i.e.,~those in value categories $V_i$ such that $V_i > V_\text{ave}$  (e.g.,~senior engineers, vice presidents, CEO), who feel that their contributions are not fairly valued and compensated for by A, will therefore be motivated to leave for other companies where they are offered higher salaries. Hence, in order to survive A will be forced to match the salaries offered by others to retain these employees, thereby forcing the distribution to spread to the right of $S_\text{ave}$, as seen in figure~\ref{spread}(b).

At the same time, the generous compensation paid to all employees in categories $V_i$ such that $V_i < V_\text{ave}$, will motivate candidates with the relevant skill sets (e.g.,~low-level administration, sales and marketing staff) from other companies to compete for these higher paying positions in A. This competition will eventually drive the compensation down for these overpaid employees forcing the distribution to spread to the left of $S_\text{ave}$, as seen in Figure~\ref{spread}(c). Eventually, we will have a distribution that is not a delta function, but a broader one where different employees earn different salaries depending on the values of their contributions. The funds for the higher salaries now paid to the formerly underpaid employees (i.e.,~those who satisfy $V_i > V_\text{ave}$) come out of the savings resulting from the reduced salaries of the formerly overpaid group (i.e.,~those who satisfy $V_i < V_\text{ave}$), thereby conserving the total salary budget $M$.

Thus, we see that concerns about {\em fairness} in pay cause the emergence of a more equitable salary distribution in a free market environment through its self-organizing, adaptive, evolutionary dynamics and that its spread is closely related to fairness in relative compensation. The point of this analysis is not to model the exact details of the free market dynamics but to show that the {\em notion of fairness} plays a central role in driving the emergence and spread of the salary (in general, utility) distribution through the free market mechanisms.

Even though an individual employee cares only about her utility and no one else's, the collective actions of all the employees, combined with the profit maximizing survival actions of all the companies, in an ideal free market environment of supply and demand for talent, under resource constraints, lead towards a more fair allocation of wages, guided by Adam Smith's ``invisible hand'' of self-organization.

We have used salary as a proxy for utility in this example to motivate the problem. In general, utility for an employee is a complicated aggregate that depends on a host of factors, some measurable, some not. Obviously, pay (i.e.,~total compensation including base salary, bonus, options etc.) is an important component of the utility. Other components include, quantity and quality of the work or effort, authority and power of the position, job security, competition, career and personal growth opportunities, work schedule, retirement and health benefits, peer appreciation and recognition, company culture and work environment, job location, and so on, not necessarily in that order.

Given this free market dynamics scenario, three important questions arise: (i) Will this self-organizing dynamics lead to an equilibrium distribution or will the distribution continually evolve without ever settling down? (ii) If there exists an equilibrium distribution, what is it? (iii) Is this distribution fair?

Our knowledge of the free market dynamics is incomplete, in an important way, without an answer to these fundamental questions. This requires a theoretical understanding of the free market dynamics, at a reasonable level of depth and detail, particularly from the bottom up, agents-based, perspective as described above. Given the obvious complexity of this dynamics, it is unrealistic to expect to develop a theory, and the associated models, that will address all the details and nuances. Therefore, our goal is to develop a theoretical, quantitative, framework that identifies the key concepts and general principles, helps us model and analyze free market dynamics under ideal conditions, and address these central questions. We propose such a framework in the following sections.

\section{A Game Theoretic Framework: ``Restless'' Agents Model}

\subsection{Formulating the payoff function}

We believe these questions can be addressed using a potential game theoretic framework. Continuing with the scenario described above, we assume that all employee agents are generally ``dissatisfied'' in their current positions, due to aforementioned unfairness considerations. In our model, every employee feels that she is {\em unfairly undervalued} compared to others in their peer group. Every employee feels they could be doing better, they should be doing better, given their talents and experience, in their company or elsewhere. As a result, they all are constantly on the lookout for job opportunities to improve their utilities. That is, these {\em utility-maximizing, fairness-seeking, teleological} agents are always {\em restless}, itching to move.

Even though the utility for an employee is a complex aggregate of several factors, we propose that it is broadly composed of three dominant elements: (i) utility derived from salary, (ii) disutility from effort, and (iii) utility from fairness. The first two are rather straightforward to see, but the third requires some more discussion along the lines of the scenario described above. The first two help us model the tendency of an employee to maximize one's utility from salary while minimizing the effort put into receiving it.

As for the third, consider the following. At any job level, an agent is looking to improve her utility only in the jobs space that is {\em accessible} to her based on her education, experience, and other such qualifications. That is, a receptionist is not eyeing the job announcement for a CEO. In that sense, what matters in trying to improve one's utility is the {\em local} competition at the agent's level. It is the assessment of one's {\em relative} status in a peer group that matters, not its absolute value. For instance, a vice president is not necessarily very happy that she is enjoying much more utility than her receptionist, but is extremely unhappy that her peer, another vice president with comparable (or perhaps even less) skills and contributions, has been better recognized in the organization with awards, better work assignments, more perks etc., thereby enjoying a higher utility than her.  As far as this ``unhappy'' agent is concerned, the metric that matters to her is whether she is {\em one of the chosen few} or {\em one of the many} in her peer level. Her preference is to be one of the few and possibly the only one enjoying a lot of utility. This is irrespective of the category one is in. The question is not about money, but about a fair valuation and recognition of one's abilities and contributions to the organization. Thus, it is a fairness issue and it matters a lot to people. This is what we attempt to capture in our formulation.

Combining all three, we have 
\begin{equation}
\mathrm{h}_i(S_i,E_i,N_i)=\mathrm{u}(S_i)-\mathrm{v}(E_i)+\mathrm{f}(N_i),\label{utility_1}
\end{equation} 
where $h_i$ is the total utility of an employee earning a salary $S_i$ by expending an effort $E_i$, while competing with ($N_i -1$) other agents in the same job category $i$ for a fair recognition of one's contributions. $\mathrm{u}(\cdot)$ is the utility derived from salary, $\mathrm{v}(\cdot)$ the disutility from effort, and $\mathrm{f}(\cdot)$ is the utility from fairness that depends only on the number of agents in any given category $i$. We propose the following functional forms for these three elements:
\begin{eqnarray}
\mathrm{u}(S_i)&=\alpha\ln S_i\label{payoff_u}\\
\mathrm{v}(E_i)&=\beta(\ln S_i)^2\label{payoff_v}\\
\mathrm{f}(N_i)&=-\gamma\ln N_i,\label{payoff_f}
\end{eqnarray}
where $\alpha,\beta,\gamma>0$. The first one is easy to see, it is the commonly used logarithmic utility function. As for the second, the effort, and hence the disutility, associated with a new job are hard to calculate accurately as there are many uncertainties such as the kind of work, work hours, office politics, company culture, relocation anxieties, and so on, which are often difficult, if not impossible, to quantify in real life. In addition, the effort term not only captures the quantity and quality of work involved, but also the necessary qualifications such as education, skills, experience, etc., needed at the job. For example, someone with no medical training, obviously, will not be able to perform the work of a cardiac surgeon successfully, no matter how hard he or she tries. Typically, one estimates and compensates for this disutility of effort by negotiating a salary package that would make it worth the effort. Thus, in practice, one intuitively uses salary as a proxy to estimate the effort and compensate for the disutility. 

We find empirical evidence that supports this line of reasoning in the work of Stratton~\cite{stratton2011why} and  Ahituv and Lerman~\cite{ahituv2007how} who have demonstrated that effort correlates with log(salary). Furthermore, our model is consistent with the conditions imposed on effort $E$ as a function of salary, $\mathrm{E}(S)$. According to Katz~\cite{katz1986efficiency} and Akerlof and Yellen~\cite{akerlof1986efficency}, $\mathrm{E}(S)$ should satisfy the following conditions: 
\begin{enumerate}[(i)]
\item $\mathrm{d}E/\mathrm{d}S > 0$
\item $\mathrm{E}(0) \leq 0$ and 
\item the elasticity $S/E\times(\mathrm{d}E/\mathrm{d}S)$ should be decreasing. 
\end{enumerate}
Our effort function $\mathrm{E}(S) = \ln S$ satisfies all three conditions:
\begin{enumerate}[(i)] 
\item $\mathrm{d}E/\mathrm{d}S = 1/S > 0$
\item $\mathrm{E}(0) = -\infty < 0$
\item $S/E\times(\mathrm{d}E/\mathrm{d}S)= (S/\ln S) \times (1/S) = 1/\ln S$ is decreasing.
\end{enumerate} 
Therefore, our formulation for effort as a function of salary, disutility as a quadratic function of effort following~\cite{cadenillas2002executive,holmstrom1991multitask,nalebuff1983prizes,laffont1988dynamics,laffont1987auctioning,zabojnik1996pay}, and hence $\mathrm{v} = \beta(\ln S)^2$ is an eminently reasonable formulation supported both by empirical evidence as well as by theoretical expectations.

For the utility derived from fairness, we believe the negative logarithmic form captures the agents' preferences and behavior correctly as explained below. Since $N_i\in[0,\infty]$, $f_i$ is $[\infty,-\infty]$. This payoff function may be intuitively interpreted as capturing the following:
\begin{enumerate}[(i)]
\item When $N_i\to\infty$, $f_i\to-\infty$, i.e.,~an agent in level $i$ feels ``unfairly'' treated,  undervalued and underrecognized as there are so many other agents at the same job category, thereby reducing her utility.
\item When $N_i = 0$, $f_i = \infty$. This is the state where the agent can be potentially the ``happiest'', most valued and appreciated, the state all the agents strive for. But since this is an elusive state (for when the agent arrives at this level, $N_i$ is no longer 0 but 1, and $f_i = 0$ and not $\infty$), the agents are constantly in motion, restless, chasing after this dream of the ``perfect'' job. While this is clearly a simplification of what really happens in the market place, we believe, this stylized model nevertheless captures an essential, and dominating, aspect of the dynamical behavior of fairness-seeking, utility-maximizing, teleological agents, namely, {\em restlessness}, that most employees feel in the real world.
\end{enumerate}

In general, $\alpha$, $\beta$ and $\gamma$, which model the relative importance an agent assigns to these three elements, can vary from agent to agent. However, for the sake of simplicity, we assume that all agents have the same preferences and hence treat these as constant parameters (However, see section~\ref{sec:bi} -- A Bi-population Game -- for a special case). Further, presumably, there are other expressions one could use to model these three elements, but the choices we have made have interesting properties, revealing important insights and connections as we shall see shortly.

In order to move to a job with better utility, an agent needs job offers. So, the employee agents constantly gather information and scout the market, and their own companies, for job openings that are commensurate with their skill sets, experiences and career and personal goals.  Similarly, the company agents (through their human resources department, for example) also conduct similar searches looking for opportunities to fire and hire employees so that their profits may be improved.

At any given time, an employee agent is faced with one of five job options: (i) no new job offer is available, (ii) new offer has the {\em same} utility as the current one, (iii) new offer has {\em less} utility than the current one, (iv)  new offer has {\em more} utility, or (v) is let go from the current job (i.e.,~{\em zero} utility). The agent's best strategies for the five options are: for (i), (ii) and (iii), the agent stays put in the current position at the current utility, for (iv) accept the new offer, and (v) leave the company and look for a new position. Each agent's strategy is independent of what the other agents are doing.

We are now ready to answer the first question.

\subsection{Is there an equilibrium distribution?}

In a potential game framework, payoff is the gradient of potential $\phi(\mathbf{x})$, i.e.,
\begin{equation}
\mathrm{h}_i(\mathbf{x})\equiv {\partial \phi(\mathbf{x})}/{\partial x_i},
\end{equation}
where $x_i=N_i/N$ and $\mathbf{x}$ is the population vector. Therefore, by integration (we replace partial derivative with total derivative because $\mathrm{h}_i(\mathbf{x})$ can be reduced to $\mathrm{h}_i(x_i)$ expressed in Equations~\eqref{utility_1}-\eqref{payoff_f}), 
\begin{eqnarray}
\phi(\mathbf{x})&=\sum_{i=1}^n\int \mathrm{h}_i(\mathbf{x})\mathrm{d}x_i,
\end{eqnarray}
we obtain the potential of the game:
\begin{equation}
\phi(\mathbf{x})=\phi_{u}+\phi_v+\phi_f+\text{constant},\label{pay_potential}
\end{equation}
where
\begin{eqnarray}
\phi_u&=\alpha\sum_{i=1}^nx_i\ln S_i\\
\phi_v&=-\beta \sum_{i=1}^nx_i(\ln S_i)^2\\
\phi_f&=\frac{\gamma}{N} \ln \frac{N!}{\prod_{j=1}^n(Nx_i)!}.\label{fair_potential}
\end{eqnarray}
We can show that $\phi(\mathbf{x})$ is strictly concave:
\begin{eqnarray}
{\partial^2 \phi(\mathbf{x})}/{\partial x_i^2}=-{\gamma}/{x_i}<0.
\end{eqnarray}
Therefore, a unique Nash Equilibrium for this game exists, where $\phi(\mathbf{x})$ is maximized, as per the well-known theorem~\cite[p. 60]{sandhom2010population}.

It is important to note that this is a stable equilibrium as long as the evolutionary dynamics satisfies positive correlation (e.g.,~replicator dynamics, Smith dynamics, best response dynamics, etc.), for the potential is a Lyapunov function under such condition, with a guarantee of global convergence~\cite[p. 223]{sandhom2010population}.

This answers our first question.

\subsection{Connection with statistical mechanics}

Readers familiar with statistical mechanics will recognize the potential component  $\phi_f$  as entropy, and that maximizing the payoff potential in game theoretic equilibrium would correspond to maximizing entropy in statistical mechanical equilibrium, revealing a deep and useful connection between these seemingly different conceptual frameworks. This connection suggests that one may view the statistical mechanics approach to molecular behavior, also called {\em statistical thermodynamics}, from a potential game perspective. In this approach, one may view the molecules as restless agents in a game (let's call it the {\em thermodynamic game}), continually jumping from one energy state to another through intermolecular collisions. However, unlike employees who are continually driven to switch jobs in search of better utilities they desire, molecules are {\em not teleological, i.e.,~not goal-driven}, in their constant search. As prisoners of Newton's Laws, constantly subjected to intermolecular collisions, their search and dynamical evolution is the result of thermal agitation.

\subsection{What is the equilibrium distribution?}

This connection to statistical thermodynamics, and the insight that $\phi_f$  is {\em entropy} in this context, helps us in answering the second question: What is the equilibrium distribution?

We first answer this question for the thermodynamic game. Approaching the thermodynamic game from potential game perspective, we have the following ``utility'' for molecules in state $i$:
\begin{equation}
\mathrm{h}_i(E_i,N_i)=-\beta E_i - \ln N_i,\label{thermo_payoff}
\end{equation}
where $E_i$ is the energy of a molecule in state $i$, $\beta = 1/kT$, $k=1.3806488\times10^{-23}$ JK$^{-1}$ is Boltzmann constant; and $T$ is temperature. By integrating the utility, we can obtain the potential of the thermodynamic game:
\begin{equation}
\phi(\mathbf{x})=-\frac{\beta}{N} E +\frac{1}{N} \ln\frac{N!}{\prod_{i=1}^n(Nx_i)!},\label{thermo_potential}
\end{equation}
where $E=N\sum_{i=1}^nx_iE_i$ is the total energy that is conserved.

We use the method of Lagrange multipliers with $L$ as the Lagrangian and $\lambda$ as the Lagrange multiplier for the constraint $\sum_{i=1}^nx_i=1$:
\begin{equation}
L=\phi+\lambda(1-\sum_{i=1}^nx_i).\label{lagrangian}
\end{equation}
Solving $\partial L/\partial x_i=0$ and substituting the results back to $\sum_{i=1}^nx_i=1$, we obtain the well-known Gibbs-Boltzmann exponential distribution at equilibrium: 
\begin{equation}
x_i=\frac{\exp(-\beta E_i)}{\sum_{j=1}^n\exp(-\beta E_j)}.\label{boltzmann}
\end{equation}
What we just now did is the standard procedure followed in maximum entropy methods in statistical mechanics and information theory to identify the distribution that maximizes entropy under the given constraints~\cite{jaynes1957information,jaynes1957information2,kapur1989maximum}. Once again, readers familiar with statistical thermodynamics will recognize that from~\eqref{thermo_potential}, we have:
\begin{equation}
\phi=-\frac{1}{NkT} (E-TS)=-\frac{\beta}{N} A,
\end{equation}
where $A$ is the Helmholtz free energy,  $S$ is entropy, and $T$ is temperature.

For the teleodynamic game, i.e.,~the pay distribution game, we carry out the same procedure to maximize $\phi(\mathbf{x})$ in Equations~\eqref{pay_potential}-\eqref{fair_potential} to obtain the following lognormal distribution at equilibrium:
\begin{equation}
x_i=\frac{1}{S_iZ}\exp\left[-\frac{\left(\ln S_i-\frac{\alpha+\gamma}{2\beta}\right)^2}{\gamma/\beta}\right]\label{logn_potential},
\end{equation}
where $Z=N\exp\left[\lambda/\gamma-(\alpha+\gamma)^2/4\beta\gamma\right]$ and $\lambda$ is the Lagrange multiplier.

\subsubsection{Replicator Dynamics}

Alternatively, we can approach this question from the replicator dynamics point of view in game theory~\cite{sandhom2010population}. In this approach,  an agent revises its strategy based on
\begin{equation}
\rho_{ij}\propto x_j[h_j-h_i]_+.\label{imitative}
\end{equation}
Under this protocol, an agent in the job category $i$ who receives a revision opportunity, i.e.,~a new job offer in category $j$,  switches from $i$ to $j$ with probability $\rho_{ij}$. Therefore the dynamics becomes:
\begin{eqnarray}
\dot{x}_i&\propto x_i(h_i-\sum_{j=1}^nx_jh_j).
\end{eqnarray}
The equilibrium is reached (i.e.,~$\dot{x}=0$) when individual payoff equals the average payoff of the system:
\begin{equation}
h^*_i = \sum_{j=1}^nx_jh^*_j=h^*.\label{h_star}
\end{equation}
We ignore the trivial solution of $x_i=0$. Substituting this equation back in our utility function (Equation~\eqref{utility_1}), we solve to find the equilibrium distribution to be
\begin{equation}
x_i=\frac{1}{S_iZ}\exp\left[-\frac{\left(\ln S_i-\frac{\alpha+\gamma}{2\beta}\right)^2}{\gamma/\beta}\right],\label{logn_replicator}
\end{equation}
where $Z=N\exp\left[h^*/\gamma-(\alpha+\gamma)^2/4\beta\gamma\right]$. This result agrees with~\eqref{logn_potential}.

This result is also in agreement with what Venkatasubramanian~\cite{venkatasubramanian2009what,venkatasubramanian2010fairness} derived using an information theoretic framework. In that approach, the constraints are determined by information typically known about the distribution {\em a priori}. They are: (i) total number of employees $N$, (ii) total amount of money $M$ budgeted to pay all these employees, (iii) minimum salary, $S_\text{min}$, received by the lowest paid employee, often fixed by the minimum wage law or a reservation wage, and (iv) the maximum salary, $S_\text{max}$, cannot exceed $M$. As Venkatasubramanian has shown, maximizing entropy under these constraints leads to a lognormal distribution at equilibrium given by:
\begin{equation}
\mathrm{f}(S;\mu,\sigma)=\frac{1}{S\sigma\sqrt{2\pi}}\exp\left[-\frac{(\ln S-\mu)^2}{2\sigma^2}\right],\label{logn_entropy}
\end{equation}
where $\mu=\ln(M/N)$; $\sigma=(\ln M-\ln S_\text{min})/2 a$; and $a$ is a parameter chosen using the Chebychev inequality given by:
\begin{equation}
\text{Prob}(-a\sigma<X-\mu<a\sigma)\geq1-\frac{1}{a^2},
\end{equation}
to the level of confidence desired in the estimate for $\sigma$ (e.g.~for $a=10$, $P\geq 0.99$). Equation~\eqref{logn_entropy} is same as~\eqref{logn_potential} or~\eqref{logn_replicator} with the following identities:
\begin{equation}
\begin{cases}
\mu &= \frac{\alpha+\gamma}{2\beta}\\
\sigma &=\left(\frac{\gamma}{2\beta}\right)^{1/2}
\end{cases}.
\end{equation}

For the thermodynamic game, it is easy to show from Equations~\eqref{imitative} through~\eqref{h_star}, and~\eqref{thermo_payoff}, a similar replicator dynamics analysis produces the same Gibbs-Boltzmann exponential distribution in~\eqref{boltzmann} at equilibrium.

Thus, we see that, intuitively, maximizing the game theoretic potential (Equation~\eqref{pay_potential} or~\eqref{thermo_potential}) is the same as maximizing entropy subject to the constraints. In the statistical mechanical or information theoretic formulations, these constraints are separately imposed on entropy  whereas in the game theoretic formulation (Equation~\eqref{pay_potential} or~\eqref{thermo_potential}) the constraints are already embedded in the equation (the only additional constraint imposed is the total number of agents, $N$). Therefore, the resulting Lagrangian (e.g.,~Equation~\eqref{lagrangian}) is the same, thereby leading to the same distribution. These demonstrate the internal consistency among the three different approaches, namely, potential game theory, replicator dynamics, and statistical mechanics, which is reassuring.

\subsubsection{A Bi-population Game}\label{sec:bi}

A system with indistinguishable and identical agents is a simplification, of course, of reality and the lognormal distribution represents the intrinsic inequality for such an ideal scenario. There are other factors that could further skew this inequality. One such factor is the heterogeneity among agents, i.e.,~different agents having different $\alpha$, $\beta$, and $\gamma$ values. As an illustrative example, we present an analysis for a bi-population system containing two classes of agents each with a distinct set of $\alpha$, $\beta$, and $\gamma$ values. The utility of an agent at salary level $i$ is therefore defined as
\begin{equation}
h_{i,j}=\alpha_j \ln S_i -\beta_j (\ln S_i)^2-\gamma_j\ln (N_{i,1}+N_{i,2}),
\end{equation}
where the choice of $j\in\{1,2\}$ indicates either Class 1 population or Class 2 population.

The equilibrium replicator dynamics is also modified:
\begin{equation}
\begin{cases}
h^*_{i,j}&=h^*_j\quad \forall i\in\Omega_j\\
h_{k,j}&<h^*_j\quad\forall k\notin\Omega_{j},
\end{cases}\label{bi_rep}
\end{equation}
where $\Omega_j = \{k|x^*_{k,j}>0\}$ denotes the collection of levels with class $j$'s presence. The first condition is identical to the homogeneous scenario. The second indicates the possibility that some levels are only occupied by a single class (i.e.,~the utility is too low for the other class).

We can prove that $\Omega_1 \cup \Omega_2 = \{k|1\leq k\leq n\}$ and $\Omega_1 \cap \Omega_2 = \emptyset$, i.e.,~every salary level contains some population but not both. First, suppose there are empty salary levels. They will soon be occupied because of the infinitely high utilities. Thus $\Omega_1$ and $\Omega_2$ cover the whole domain. Second, suppose there is an overlap where $\hat{\Omega} = \Omega_1 \cap \Omega_2$. Let the equilibrium population density be $x^*$. Equation~\eqref{bi_rep} can be rewritten as:
\begin{equation}
h_j^*=\alpha_j\ln S_i -\beta_j(\ln S_i)^2-\gamma_j\ln Nx^*\quad\forall i\in\hat{\Omega}.
\end{equation}
Thus
\begin{equation}
\lim_{\Delta S\to 0}\frac{\Delta x^*}{\Delta S_i}=\frac{\mathrm{d}x^*}{\mathrm{d}S}=\frac{\alpha_j-2\beta_j\ln S}{\gamma_jS}x^*.
\end{equation}
This indicates two distinct gradients for {\em every} point in $\hat{\Omega}$. Therefore we will not see an overlapping region with mixed populations.

We can also prove that the equilibrium density curve is continuous at the interface of two populations. Suppose otherwise, at the interface $S=\hat{S}$,
\begin{align}
x^*_1(\hat{S})\neq x^*_{2}(\hat{S})
\end{align}   
according to Equation~\eqref{bi_rep} again,
\begin{equation}
\begin{split}
&\alpha_j\ln \hat{S} -\beta_j(\ln \hat{S})^2 -\gamma_j\ln x^*_j\\
\geq &\alpha_{j}\ln \hat{S} -\beta_{j}(\ln \hat{S})^2 -\gamma_{j}\ln x^*_{-j}
\end{split},
\end{equation}
i.e.,
\begin{equation}
x^*_j\geq x^*_{-j}.
\end{equation}
The only possible solution is $x^*_1=x^*_2$ therefore the population density is continuous at the interface.

Even though we now know these equilibrium characteristics of a bi-population game, an exact equilibrium density is still tedious to obtain unless $\Omega_1$ and $\Omega_2$ are given:
\begin{equation}
\begin{split}
x_i&=\frac{N_1/N}{S_iZ_1}\exp\left[-\frac{\left(\ln S_i - \frac{\alpha_1+\gamma_1}{\beta_1}\right)^2}{\gamma_1/\beta_1}\right]\mathbbm{1}(i\in\Omega_1)\\
&+\frac{N_2/N}{S_iZ_2}\exp\left[-\frac{\left(\ln S_i - \frac{\alpha_2+\gamma_2}{\beta_2}\right)^2}{\gamma_2/\beta_2}\right]\mathbbm{1}(i\in\Omega_2)
\end{split}
\end{equation}
where $Z_j$ is the normalization that ensures $\sum_{i=1}^nx_i=1$. We can, however, get a good approximation when the two lognormal curves of Class 1 and Class 2 are sufficiently separated such that the overlap is insignificant.  The overall distribution is then estimated as a mixture of two lognormal distributions: 
\begin{equation}
\begin{split}
x_i&\approx\frac{N_1}{S_iZ}\exp\left[-\frac{\left(\ln S_i - \frac{\alpha_1+\gamma_1}{\beta_1}\right)^2}{\gamma_1/\beta_1}\right]\\
&+\frac{N_2}{S_iZ}\exp\left[-\frac{\left(\ln S_i - \frac{\alpha_2+\gamma_2}{\beta_2}\right)^2}{\gamma_2/\beta_2}\right]
\end{split}\label{eqn:combined}
\end{equation}
where $N_j$ denotes the number of class $j$ agents and $Z$ denotes the normalization parameter which is easily computed.

\begin{figure*}[!]
\centerline{\includegraphics[width=0.8\linewidth]{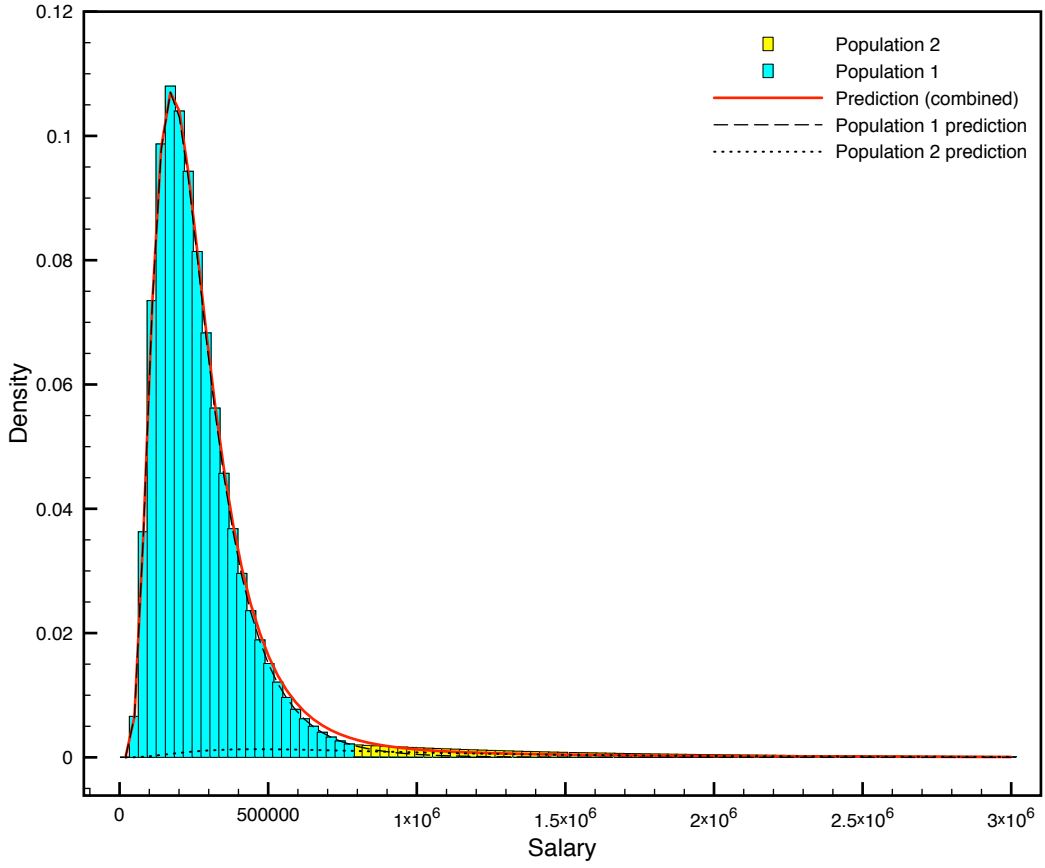}}
\caption{Simulation results (the subview features salaries from 0.5 to 3 million)}\label{bi-plot}
\end{figure*}

To test the model predictions, we ran an agent-based simulation comprising of  one million agents in two classes, at 100 salary levels, with a minimum pay of \$20,000 (using a minimum wage of \$10/hr and 2000 hours/year) and a maximum pay of \$3,000,000. We explored the typical case of 95\% of the population in Class 1 and 5\% in Class 2. The respective $\alpha$, $\beta$, and $\gamma$ values for the two classes are shown below. The dynamics unfolds by having each agent trying to maximize its utility (given by Equations~\eqref{utility_1} to \eqref{payoff_f}) by switching from its current job to a better one and the equilibrium (stationary) distribution emerges over time, as shown in figure~\ref{bi-plot}.

\begin{table}[h]
\begin{center}
\begin{tabular}{c|ccc}
\hline
$j$ &$\alpha_j$&$\beta_j$&$\gamma_j$\\
\hline
1&215&20.5&5\\
2&220.5&19.45&10
\end{tabular}
\end{center}
\label{table1}
\end{table}

\begin{figure}[!]
\centerline{\includegraphics[width=1\linewidth]{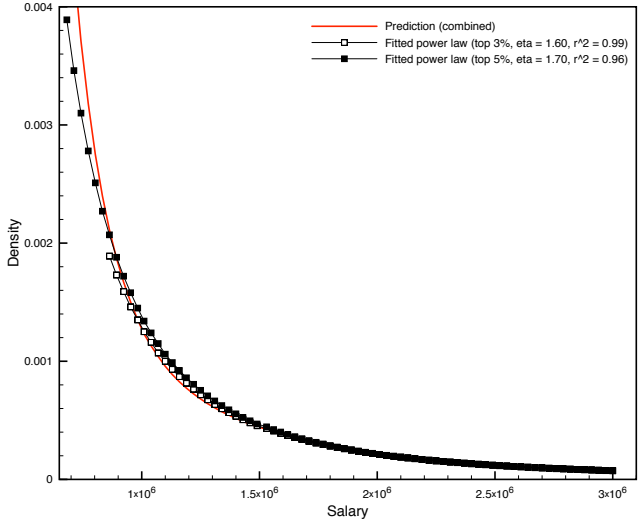}}
\caption{Fitted power law}\label{bi-plot-sub}
\end{figure}

In the figure~\ref{bi-plot}, the blue (Class 1) and yellow (Class 2) histogram bars are data from the simulation and the lines are predictions by the model. As the results show, the two populations are sufficiently separated and hence the individual lognormal distributions predicted by the model (Equation~\eqref{eqn:combined}) fit the data very well. For the population shown in yellow (Class 2), its higher $\alpha$, lower $\beta$,  and higher $\gamma$ make the utilities from salary and fairness components high enough to more than compensate for the disutility from effort, that the agents from this population are strongly motivated to jobs with higher salaries. They are also averse to jobs with lower pay. It is the opposite for the agents from population 1. Recall that by effort we don't mean just the effort expended in carrying out a job, but we also include all the effort one has invested over the years in acquiring the necessary education, skills and experience to be able to perform the job. For the agents from the yellow population (Class 2), given their $\alpha$, $\beta$, and $\gamma$ values, it's worth their effort and thus they end up occupying these higher paying jobs. We observe that the combined distribution (solid red line), as one might expect, fits the lognormal distribution for the blue population (Class 1) quite well in the lower and medium salary ranges but deviates from it for higher salaries.

We also show that the distribution of the higher salaries (largely occupied by the yellow population agents) can be fitted to an inverse power law, given as follows:
\begin{equation}
x_i\propto S_i^{-(1+\eta)}
\end{equation}
\begin{enumerate}[(i)]
\item Top 3\% fitted: $\eta=1.60$, $r^2=0.99$
\item Top 5\% fitted: $\eta=1.70$, $r^2=0.96$
\end{enumerate}

We see that the inverse power law fit is very good for both top 3\% and 5\%. The Pareto exponents from our simulation data agree quite well with empirical data reported in the literature --  between 1 and 2, but typically around 1.5 for the top 3\% \cite{chakrabarti2013econophysics}. Thus, the main lesson here is that  while the overall distribution is a combination of two lognormal distributions, it can be quite easily misidentified as a lognormal for the majority and an inverse power law or Pareto distribution for the minority at the top end of the salaries. This again confirms similar warnings by Perline~\cite{perline2005strong} and Mitzenmacher~\cite{mitzenmacher2004brief}. For actual salary distributions reported in the literature~\cite{chakrabarti2013econophysics}, the available data is not good enough to sort this out clearly and further studies are needed. 
 Our 2-class approach can be generalized for the $n$-class game along the lines we described above. However, in practice, we suspect that we might only need four classes -- low income (``blue collar'' jobs), middle income (mid-level ``white collar'' jobs), upper income (upper management jobs) and extreme income (c-suite executives) --  to  model income distribution data adequately. At any rate, at the present time, the empirical data reported in the literature is not good enough to test 3-class or 4-class models. The best it seems to be able to do is to identify the need for a 2-class model, but even there it is unable to discriminate between a lognormal distribution and a power law fit for the top 3-5\% as we showed above.

An interesting question one might ask is ``Why does the 2-class split in actual data occur at about 95-97\% of the population? Why not at 80\%, for instance?'' In our theory, this is related to the fraction of the population that is highly motivated, talented, and driven towards individual accomplishments and success (i.e., Class 2).  It would be nice if we had data that directly showed where this 2-class split occurs in the real world, but we donÕt. One approximation we can perhaps use on how human abilities are distributed in a population is how IQ is distributed in a population. Obviously, as we all know, IQ doesnÕt capture the complete picture of the human talent spectrum and how people succeed. Nevertheless, it is interesting to note that 2-$\sigma$ deviation (IQ = $\sim$130) from the median IQ value occurs at $\sim$97\% of the population -- i.e., the top $\sim$3\% of the population have an IQ greater than $\sim$130, beyond the 2-$\sigma$ deviation.

\section{Discussion and Conclusion}

There has been some work in the past that explored the connection between game theory and statistical mechanics~\cite{blume1993statistical,wolpert2006information,kikkawa2009statistical}. What is new about our contribution is that it shows a direct and deep connection between the dynamics of animate, fairness-driven, utility-maximizing, rational {\em teleological} agents and inanimate, purpose-free, thermally-driven molecular entities. Our result reveals the surprising and important connection between entropy and game theoretic potential, demonstrating that the statistical thermodynamic equilibrium reached by molecules is really a Nash Equilibrium. We believe that this is a significant insight, for it suggests that statistical thermodynamics can be seen as a special case of potential game theory. Alternatively, one may view this insight as the generalization of the laws of statistical thermodynamics to teleological systems, such as economic systems, yielding a new conceptual framework, which we call {\em statistical teleodynamics}, that unifies statistical thermodynamics and population game theory. This framework bridges the conceptual gulf mentioned in the introduction, as our ideal teleological agents are rational, fairness-seeking, utility maximizing strategists, with a natural connection to statistical thermodynamics.

As noted, one could presumably choose other expressions to model the three elements (Equations~\eqref{payoff_u}-\eqref{payoff_f}) in~\eqref{utility_1}, but it is not clear whether they will necessarily lead to the Gibbs-Boltzmann distribution, Helmholtz free energy and entropy in the limiting case of the thermodynamic game involving molecules. We find this correspondence to be particularly appealing, in fact comforting, that statistical teleodynamics properly reduces to well-known results in statistical thermodynamics as a limiting case. This universality has a nice ring to it.

Another important observation is that, in statistical thermodynamics, the claim about the equilibrium state is a {\em probabilistic} one~--~it is the {\em most probable} outcome, one where entropy is maximum. However, our game theoretic result shows that the Nash Equilibrium state reached by the molecules, the one that maximizes the potential $\phi(\mathbf{x})$,  is a {\em deterministic} outcome, not a probabilistic one. This observation has potentially important implications concerning the philosophical foundations of statistical thermodynamics, and that of information theory, such as ergodicity and metric transitivity~\cite{tolman1938principles,reif1965fundamentals,nash2012elements,jaynes1957information,jaynes1957information2,khinchin1957mathematical,theil1967economics,sklar1998physics}, but we are not addressing them here.

As we have shown, the deep connection between game theory and statistical mechanics, and with information theory, occurs via entropy, a concept that is often misunderstood and much maligned~\cite{venkatasubramanian2009what,venkatasubramanian2010fairness,samuelson1990gibbs,sen1997economic}. In the past, there have been several attempts to find a suitable interpretation of entropy for economic systems without much success~\cite{georgescu1986entropy,proops2004modelling,richmond2013econophysics,samuelson1990gibbs}.   In these attempts, one typically wrote down equations in economics that mimicked expressions in thermodynamics for entropy, energy, temperature, etc. -- but no identification of entropy in terms of meaningful economic concepts was made. Just as entropy is a measure of disorder in thermodynamics and uncertainty in information theory, what does entropy mean in economics? Neither interpretation, disorder nor uncertainty, makes much sense in the economic context. Economic systems work best when they have orderly markets. Why then would anyone want to maximize disorder? Similarly, economic systems work best when there is less uncertainty. Why then would anyone want to maximize uncertainty? The inability to resolve this crucial issue has been a major conceptual roadblock for decades thwarting meaningful progress, as evidenced from Amartya Sen's remarks about the Theil index~\cite{sen1997economic} or Paul Samuelson's objections to entropy in economics~\cite{samuelson1990gibbs}.

The crucial insight here is the recognition that entropy is a measure of {\em fairness} in a distribution, an insight that has not been explicitly recognized and particularly stressed in prior work in statistical thermodynamics, information theory, or economics. Despite the several attempts in the past, entropy has played, by and large, only a marginal role in economics, even that with strong objections from leading practitioners. Its pivotal role in economics and in free market dynamics has never been recognized. This is mainly because entropy's essence as fairness appears as different facets in different contexts~\cite{venkatasubramanian2010fairness}. In thermodynamics, being {\em fair} to all accessible phase space cells at equilibrium under the given constraints~--~i.e.,~assigning {\em equal} probabilities to all the allowed microstates~--~projects entropy as a measure of {\em randomness or disorder}~\cite{reif1965fundamentals}. This is the appropriate interpretation in this particular context, but it obscures the essential meaning of entropy as a measure of fairness. In information theory, being {\em fair} to all messages that could potentially be transmitted in a communication channel~--~i.e.,~assigning equal probabilities to all the messages~--~shows entropy as a measure of {\em uncertainty}~\cite{jaynes1957information,jaynes1957information2}. Again, while this is the appropriate interpretation for this application, this, too, conceals the real nature of entropy. In the design of teleological systems, being fair to all potential operating environments, entropy emerges as a measure of {\em robustness} i.e.,~maximizing system safety or minimizing risk~\cite{venkatasubramanian2007theory}. Once again, this is the right interpretation for this domain, but this also hides its true meaning.

Thus, the common theme across all these different contexts is the essence of entropy as a measure of fairness, which stems from the notion of {\em equality} expressed mathematically. If there are $N$ possible candidates among whom a resource is to be distributed, and if no particular candidate is to be preferred over another, then the fairest distribution of the resource is one of {\em equal} allocation among all of them. This quantitative mathematical relationship is at the core of the concept of fairness. Bernoulli and Laplace expressed this notion in probability theory as the {\em Principle of Insufficient Reason}. The generalization of this principle is the {\em Principle of Maximum Entropy}~\cite{jaynes1957information} which addresses the question: ``What is the fairest assignment of probabilities of several alternatives given a set of constraints?'' Thus, the roots of entropy as a fairness measure can be traced all the way back to the  {\em Principle of Insufficient Reason}~\cite{venkatasubramanian2010fairness}. Somehow, this important insight seems to have been missed in all these years since the discovery of entropy.

It is a historical accident that the concept of entropy was discovered in the context of thermodynamics and, therefore, unfortunately, got tainted with the negative notions of doom and gloom, while, ironically, it is really a measure of fairness, which is a good thing. Even its subsequent ``rediscovery'' by Shannon in the context of information theory did not help much, as entropy now got associated with uncertainty, again not a good thing. It is important not to confuse entropy as a concept from physics even though it was discovered there. In other words, it is not like energy or momentum, which are physics-based concepts. Entropy really is a concept in probability and statistics, an important property of distributions, whose application has been found to be useful in physics and information theory. In this regard, it is more like variance which is a property of distributions, a statistical property, with applications in a wide variety of domains. However, as a result of this profound, but understandable, confusion about entropy as a physical principle, one got trapped in the popular notions of entropy as randomness, disorder, doom or uncertainty, which has prevented people from seeing the deep and intimate connection between statistical theories of inanimate systems composed of non-rational entities (e.g.,~gas molecules in thermodynamics) and of animate, teleological, systems of rational agents seen in biology, economics, and sociology.

In addition, and most crucially for economics, entropy's connection with the self-organizing free market dynamics has not been made before. Our contribution demonstrates that the ideal free market for labor promotes fairness as an emergent self-organized property and identifies entropy as the appropriate measure of this fairness. We believe that by properly recognizing entropy as a measure of fairness, a fundamental economic and social principle, and showing how it is naturally and intimately connected to the dynamics of the free market, our theory makes a significant conceptual advance in revealing the deep and direct connections between game theory, statistical thermodynamics, information theory, and economics.

As noted in the introduction, researchers in the econophysics community have proposed thermodynamical models for the emergence of income and wealth distributions~\cite{chatterjee2005econophysics,richmond2013econophysics,chakrabarti2013econophysics,willis2011wealth,willis2004evidence,richmond2006review,yakovenko2009colloquium,yakovenko2012statistical}. Even though our contribution also utilizes concepts from statistical mechanics, it takes an entirely different perspective by addressing the fairness issue. The fairness question has not been addressed in the past econophysics approaches. On the other hand, there has been a great amount of work by economists on fairness but these approaches have not addressed whether the free market dynamics will lead to a fair distribution. Indeed, the conventional wisdom in economics is that the free market for labor cares only about efficiency and not fairness. Thus, there is a disconnect between the econophysics and mainstream economics communities in this context. The former has proposed models inspired by statistical mechanical analogues but has not interpreted entropy in economically relevant terms -- in particular, it has not addressed the issue of fairness in its theories. In contrast, the latter, which has proposed many theories of fairness, has not recognized the relevance of, and connected with, the statistical thermodynamic theories. Our contribution is to identify the deep connections between these two as well as with game theory, thereby integrating the apparently disparate approaches into a unified conceptual framework.

This revelation of entropy's true meaning also sheds new light on a decades-old fundamental question in economics, as Samuelson~\cite{samuelson1972maximum} posed in his Nobel Lecture, ``what it is that Adam Smith's `invisible hand' is supposed to be maximizing'', or as Monderer and Shapley~\cite{monderer1996potential} stated regarding the potential function $P^*$ in game theory, ``This raises the natural question about the economic content (or interpretation) of $P^*$: What do the firms try to jointly maximize? We do not have an answer to this question.''

Our theory suggests that what all the agents in a free market environment are jointly maximizing, i.e.,~what the ``invisible hand'' is maximizing, is {\em fairness}. Maximizing entropy, or game theoretic potential, is the same as maximizing fairness {\em collectively} in economic systems, i.e.,~being fair to everyone under the given constraints. In other words, economic equilibrium is reached when every agent feels she or he has been fairly compensated for her or his efforts. As we all know, fairness is a fundamental economic principle that lies at the foundation of the free market system. It is so vital to the proper functioning of the markets that we have regulations and watchdog agencies that breakup and punish unfair practices such as monopolies, collusion, and insider trading. Thus, it is eminently reasonable, indeed particularly reassuring, to find that maximizing fairness collectively, i.e.,~maximizing entropy, is the condition for achieving economic equilibrium. We call this result the {\em fair market hypothesis}. We claim that the ideal free market, in addition to being efficient, also promotes fairness to the maximum level allowed by the constraints imposed on it.   A related interpretation is that the game theoretic potential captures the trade offs among utility from salary, disutility from effort and utility from fairness, for all the agents collectively. The ideal free market tries to accommodate every agent's individual preference regarding this trade off, given the overall constraints on money and job openings. Thus, in a sense, the market is trying to maximize ``harmony'', an accord freely and jointly agreed to by all the agents, where every agent feels fairly compensated for his effort.

A key prediction of this theory is that the lognormal distribution is the fairest inequality of pay in organizations under ideal conditions at equilibrium, if the employee population is homogenous. For a two-class population, we predict that the overall distribution is a combination of two different lognormal distributions corresponding to the two classes. In this case, as we showed, the upper-tail (comprising of the top earners, top $\sim$3-5\%), though lognormal, can be fitted quite accurately as an inverse power law or a Pareto distribution. It has been empirically observed that the income distribution, across different countries and different time periods, follows a lognormal distribution for the bottom $\sim$95\% and an inverse power law (Pareto) for the top $\sim$3-5\%. The empirical data supports our prediction, but we would like to observe that the Pareto fit for the top end is a mistaken identity of the underlying lognormal distribution that is different from the one that fits the bottom $\sim$95\%. Further studies are needed to carefully analyze the data to settle this issue.

Typically, econophysicists~\cite{chatterjee2005econophysics,yakovenko2009colloquium,yakovenko2012statistical} like to claim that the bottom $\sim$95\% follows Boltzmann-Gibbs (BG) exponential distribution or a gamma distribution, not lognormal, and that the top $\sim$3-5\% follows a Pareto distribution.  We beg to differ on both counts as we have shown in this paper. One main difficulty with the BG exponential or the gamma distribution claim is the interpretation of the underlying economic notions. For example, from the maximum entropy procedure which underlie these claims, we can show that the BG exponential distribution implies a utility function that is linear in salary~\cite{kapur1989maximum} which conflicts with the principle of diminishing marginal utility, one of the founding concepts of economic theory. At the risk of repeating ourselves, we emphasize that in our framework we have tried to formulate an approach that is sensible from an economic perspective -- e.g., recognizing entropy as fairness, modeling agents with reasonable utility preferences, rational agents making moves motivated by utility maximization and not due to random events, etc.

Given that different employees in an organization have different talents thereby making different contributions, some more some less, we expect them to be compensated differently. So, we naturally expect an unequal distribution of pay in an organization. This is only fair as people who contribute more should be paid more. But how much more? What is the fairest distribution of pay? In other words, what is the fairest inequality of pay? This is at the heart of the inequality debate. Our theory suggests that the lognormal distribution is the fairest inequality of pay for a homogenous population. One may view our result as an `economic law' in the statistical thermodynamics sense. The ideal free market, guided by the ``invisible hand'', will self-organize to ``discover'' and obey this economic law if allowed to function freely without collusion like practices or other such unfair interferences. This result is the economic equivalent of the Gibbs-Boltzmann exponential distribution in thermodynamics.

There are obvious limitations to our model -- we have assumed perfectly rational agents, no externalities, ideal free market conditions, and so on, which are clearly not valid in real life. However, our objective was to develop a general game theoretic framework, identify key principles, and make predictions that are not restricted by market specific details and nuances. Nevertheless, despite such simplifying assumptions, it is encouraging that our predictions are supported by empirical data. Clearly, the next steps are to conduct more comprehensive studies of pay distributions in various organizations in order to understand in greater detail the deviations from ideality in the market place. Agencies such as the Bureau of Labor Statistics and National Bureau of Economic Research could organize task forces to gather pay data from various companies and organizations. The data should be so grouped to analyze pay distribution patterns across several dimensions such as: (i) organization size -- small, medium, large, and very large number of employees, (ii) different industrial sectors, (iii) different types such as private corporations, governments (state and federal), non-profit organizations, etc. Similar studies should be conducted in other countries as well so that we can better understand global patterns.

Further work is also needed to examine whether there are other payoff functions which can explain and predict better than what we have proposed. Another obvious line of research is combining this approach with behavioral models., by endowing the ideal agents with non-idealities such as trending and copying behaviors.

\begin{acknowledgments}
Venkat Venkatasubramanian dedicates this contribution, with profound gratitude, to Prof. Edwin T. Jaynes, in memoriam, whose pioneering research in probability theory, statistical mechanics, and information theory inspired this work.
\end{acknowledgments}

\end{document}